\documentclass[12pt,a4paper]{article}

\usepackage{epsfig}
\usepackage{psfig}
\usepackage{exscale}
\def\lsim{\raise0.3ex\hbox{$<$\kern-0.75em\raise-1.1ex\hbox{$\sim$}}}
\def\gsim{\raise0.3ex\hbox{$>$\kern-0.75em\raise-1.1ex\hbox{$\sim$}}}
\newcommand{\<}{\langle}
\renewcommand{\>}{\rangle}
\newcommand{\be}{\begin{equation}}
\newcommand{\ee}{\end{equation}}
\newcommand{\ba}{\begin{eqnarray}}
\newcommand{\ea}{\end{eqnarray}}

\def\spose#1{\hbox to 0pt{#1\hss}}
\def\ltapprox{\mathrel{\spose{\lower 3pt\hbox{$\mathchar"218$}}
 \raise 2.0pt\hbox{$\mathchar"13C$}}}
\def\gtapprox{\mathrel{\spose{\lower 3pt\hbox{$\mathchar"218$}}
 \raise 2.0pt\hbox{$\mathchar"13E$}}}
\def\phv{\vec \phi}

\def\NT{N_\tau}
\def\nt{\ifmmode\NT\else$\NT$\fi}
\def\NS{N_\sigma}
\def\ns{\ifmmode\NS\else$\NS$\fi}

\def\PRep{{ Phys.\ Rep.\ }}

\def\p{^\prime}

\def\n{\noindent}

\begin{document}
\begin{titlepage} 
\thispagestyle{empty}

 \mbox{} \hfill BI-TP 2003/19\\
 \mbox{} \hfill Oktober 2003
\begin{center}
\vspace*{0.8cm}
{{\Large \bf Correlation lengths and scaling functions\\
 in the three-dimensional $O(4)$ model\\}}\vspace*{1.0cm}
{\large J. Engels, L. Fromme and M. Seniuch}\\ \vspace*{0.8cm}
\centerline {{\em Fakult\"at f\"ur Physik, 
    Universit\"at Bielefeld, D-33615 Bielefeld, Germany}} \vspace*{0.4cm}
\protect\date \\ \vspace*{0.9cm}
{\bf   Abstract   \\ } \end{center} \indent
We investigate numerically the transverse and longitudinal 
correlation lengths of the three-dimensional $O(4)$ model
as a function of the external field $H$. From our data we
calculate the scaling function of the transverse correlation
length, and that of the longitudinal correlation length for 
$T>T_c$. We show that the scaling functions do not only describe
the critical behaviours of the correlation lengths but encompass
as well the predicted Goldstone effects, in particular the
$H^{-1/2}$-dependence of the transverse correlation length
for $T<T_c$. In addition, we determine the critical exponent
$\delta=4.824(9)$ and several critical amplitudes from which we
derive the universal amplitude ratios $R_{\chi}=1.084(18),~
Q_c=0.431(9),~Q_2^T=4.91(8),~Q_2^L=1.265(24)$ and $U_{\xi}^c=1.99(1)$.
The last result supports a relation between the longitudinal and 
transverse correlation functions, which was conjectured to hold below
$T_c$ but seems to be valid also at $T_c$.

\vfill \begin{flushleft} 
\noindent{\rule[-.3cm]{5cm}{.02cm}} \\
\vspace*{0.2cm}
E-mail addresses: engels@physik.uni-bielefeld.de, 
fromme@physik.uni-bielefeld.de,\\ 
seniuch@physik.uni-bielefeld.de
\end{flushleft} 
\end{titlepage}


\section{Introduction}

In $O(N)$ spin models with $N>1$ two types of correlation lengths appear.
They govern the exponential decay of the correlation functions of the
transverse and longitudinal spin components, defined relative to the external
field $\vec H$. Like in the case of the magnetization and the susceptibilities,
the behaviour
of the correlation lengths in the critical region is described by asymptotic
scaling functions, critical exponents and amplitudes, which characterise the
underlying universality class. These quantities are of general interest. In
addition, there are predictions \cite{Fisher:1985,Patashinskii:1973} for the
correlation lengths, which are related to the presence of massless Goldstone
modes \cite{Zinn-Justin:1996,Anishetty:1995kj} and are still untested.
The measurement of the correlation lengths as functions of the field $H$
enables us to verify these predictions and to determine the critical
parameters and scaling functions. We have chosen the three-dimensional $O(4)$
model to carry out this program for the following reasons. Firstly, this 
model is of importance for quantum chromodynamics (QCD) with two degenerate
light-quark flavours at finite temperature, because it is believed 
\cite{Pisarski:ms}-\cite{Toldin:2003hq} to belong to the same universality 
class as QCD at its chiral transition in the continuum limit. Secondly, the
$H$-dependent model has already been investigated in some detail by Monte
Carlo methods, in particular the critical behaviours of the magnetization
\cite{Toussaint:1996qr,Engels:1999wf} and susceptibilities. In addition the 
corresponding Goldstone-mode effects have been verified \cite{Engels:1999wf}.
  
The specific model which we study here is the standard $O(4)$-invariant
nonlinear $\sigma$-model, which is defined by
\be
\beta\,{\cal H}\;=\;-J \,\sum_{<{\vec x},{\vec y}>}\phv_{\vec x}\cdot
\phv_{\vec y} \;-\; {\vec H}\cdot\,\sum_{{\vec x}} \phv_{\vec x} \;,
\ee
where ${\vec x}$ and ${\vec y}$ are nearest-neighbour sites on a 
three-dimensional hypercubic lattice, and $\phv_{\vec x}$ is a
four-component unit vector at site ${\vec x}$. It is convenient to decompose 
the spin vector $\phv_{\vec x}$ into longitudinal (parallel to the magnetic 
field $\vec H$) and transverse components 
\be
\phv_{\vec x}\; =\; \phi_{\vec x}^{\parallel} \vec e_H +
\phv_{\vec x}^{\perp} ~,\quad {\rm with}\quad \vec e_H=\vec H/ H~.
\ee
The order parameter of the system, the magnetization $M$, is then the 
expectation value of the lattice average $\phi^{\parallel}$
of the longitudinal spin components
\be
M \;=\;\<\: \frac{1}{V}\sum_{{\vec x}} \phi_{\vec x}^{\parallel} \:\>\; 
=\;\<\, \phi^{\parallel} \,\>~.
\ee
Here $V=L^3$ and $L$ is the number of lattice points per direction.
There are two types of susceptibilities. The longitudinal 
susceptibility is the usual derivative of the magnetization, 
whereas the transverse susceptibility corresponds to the fluctuation 
per component of the lattice average $\phv^{\perp}$ of the
transverse spin components
\ba
\chi_L\!\! &\!=\!&\!\! {\partial M \over \partial H}
 \;=\; V(\<\, \phi^{\parallel2} \,\>-M^2)~, \label{chil}\\
\chi_T\!\! &\!=\!&\!\! {V \over 3} \<\, \phv^{\perp2} \,\> 
\;=\;{M \over H}~. \label{chit}
\ea
The expectation value $\<\,\phv^{\perp}\,\>$ is of course zero.
The connected two-point correlation functions of the longitudinal and 
transverse spins are defined by 
\ba
G_L({\vec x})\!\! &\!=\!&\!\! \<\: \phi^{\parallel}_{\vec x}
 \phi^{\parallel}_0 \:\>-M^2~, \label{gpl}\\
G_T({\vec x})\!\! &\!=\!&\!\! {1 \over 3} \<\: \phv^{\perp}_{\vec x} \cdot
\phv^{\perp}_0 \:\>~. \label{gpt}
\ea
They are related to the susceptibilities by
\be
\chi_{L,T}\;=\; \sum_{{\vec x}} G_{L,T}({\vec x})~. \label{fludiss}
\ee
For all temperatures $T$ (the coupling $J$ acts here as inverse 
temperature, i.\ e.\ $J=1/T$) and fields $H$, except on the coexistence
line $H=0,T<T_c$ and at the critical point, the large distance behaviour
of these correlation functions is determined by the respective
exponential correlation lengths $\xi_{L,T}$
\be
G_{L,T}({\vec x}) \sim \exp (-|\vec x|/\xi_{L,T})~.
\label{xiexp}
\ee 
On the coexistence line, where the correlation functions decay according
to a power law, it is still possible to define a transverse correlation 
length \cite{Privman:1991} from the so-called stiffness constant. We do
however not consider this option here.

\subsection{Goldstone-Mode Effects}
\label{section:Gold}

The spontaneous breaking of the rotational symmetry for temperatures 
below the critical point gives rise to the so-called spin waves:
slowly varying long-wavelength spin configurations, whose energies may
be arbitrarily close to the ground-state energy - the massless Goldstone
modes \cite{Vaks:1968}. For $H\rightarrow 0$ the magnetization
$M$ below $T_c$ attains a finite value, the spontaneous magnetization
$M(T,0)$ (here and in the following we assume always that $H\ge0$, so
that $M(T,0)>0$). The transverse susceptibility $\chi_T=M/H$, which is
directly related to the fluctuation of the Goldstone modes, diverges
therefore as $H^{-1}$ when $H\rightarrow 0$ for all $T<T_c$. It is
non-trivial that also the longitudinal susceptibility $\chi_L$ is
diverging on the coexistence curve. The predicted divergence in three
dimensions is \cite{Wallace:1975}
\be
\chi_L (T<T_c,H) \sim H^{-1/2}~.
\label{chidiv}
\ee
From a phenomenological spin wave analysis and the behaviour of the
transverse susceptibility $\chi_T$ Fisher and Privman \cite{Fisher:1985}
arrived at the conclusion that the bulk correlation length, which they
identify with the transverse correlation length, diverges also when
$H\rightarrow 0$ as
\be
\xi_T (T<T_c,H) \sim H^{-1/2}~.
\label{xtdiv} 
\ee
In fact, this connection between the behaviours of the correlation
length and the respective susceptibility, namely
\be
\xi_T^2 \sim \chi_T,
\label{xichi} 
\ee
is what one usually expects \cite{Engels:2002fi}. For instance, when 
masses are defined from the corresponding inverse susceptibilities,
\be
m^2 = \chi^{-1}~,
\label{massdef} 
\ee
one would anticipate that $m \sim \xi^{-1}$. In QCD, the transverse mass
corresponds \cite{Rajagopal:1992qz} to the pion mass, $m_T = m_{\pi}$,
and the longitudinal one to the sigma mass, $m_L = m_{\sigma}$.
According to Ref.\ \cite{Fisher:1985}, the relation equivalent to Eq.\ 
(\ref{xichi}), $\xi_L^2 \sim \chi_L$, does not hold for the longitudinal
fluctuations at long wavelengths, because they are driven by the
transverse fluctuations and the bulk correlation length also sets the 
scale of the decay of $G_L({\vec x})$. This statement is substantiated 
by the well-known relation \cite{Fisher:1985,Patashinskii:1973}
between the longitudinal and transverse correlation functions at zero
field, $T<T_c$ and large distances $|{\vec x}|$
\be
G_L({\vec x}) \approx {1 \over 2} (N-1) \left[ {G_T({\vec x}) \over 
M}\right]^2~,
\label{corrcon} 
\ee
where in our case $N=4$. One expects that the relation is still valid
for small non-zero fields $H$ near the phase boundary in the region of 
exponential decay. That implies a factor of 2 between the correlation
lengths. 

The rest of the paper is organized as follows. First we discuss the
critical behaviours of the observables and the universal scaling 
functions, which we want to calculate. In Section 3 we describe 
some details of our simulations and the way we determine the correlation
lengths. Section 4 serves to find the critical amplitudes, which are
needed for the normalizations and the universal ratios. In the 
following Section 5 we discuss the scaling functions which we obtain
from our data. The Goldstone effect on the transverse correlation 
length is demonstrated in Section 6. Subsequently we investigate
the $H$-dependence of the correlation lengths in the high temperature
phase. We close with a summary and the conclusions.


\section{Critical Behaviour and Scaling Functions}
\label{section:Criti}


In the thermodynamic limit ($V\rightarrow \infty$) the observables show
power law behaviour close to $T_c$. It is described by critical amplitudes
and exponents of the reduced temperature $t=(T-T_c)/T_c$. The scaling laws
at $H=0$ are for

\n the magnetization 
\be
 M  \;=\; B (-t)^{\beta} \quad {\rm for~} t<0~,
\label{mcr}
\ee
the longitudinal susceptibility
\be
 \chi_L \;=\; C^{+} t^{-\gamma} \quad {\rm for~} t > 0~,
\label{chicr}
\ee 
and since for $H=0,t>0$ the correlation lengths coincide $\xi_T=\xi_L=\xi$ 
(like the susceptibilities)
\be
 \xi \;=\; \xi^{+} t^{-\nu} \quad {\rm for~} t > 0~.
\label{xicr}
\ee
On the critical line $T=T_c$ or $t=0$ we have for $H>0$ 
the scaling laws
\be
M \;=\; B^cH^{1/\delta} \quad {\rm or}\quad H \;=\;D_c M^{\delta}~,
\label{mcrh}
\ee
\be
\chi_L \;=\; C^cH^{1/\delta-1} \quad {\rm with}\quad C^c \;=\;B^c/\delta~,
\label{chicrh}
\ee
and for the correlation lengths 
\be
\xi_{L,T} \;=\; \xi^c_{L,T} H^{-\nu_c}~,\quad \nu_c\;=\; \nu /\beta\delta~.
\label{xicrh}
\ee
We assume the following hyperscaling relations among the
critical exponents to be valid
\be
\gamma \;=\; \beta (\delta -1), \quad
d\nu \;=\; \beta (1 +\delta), \quad 2-\eta \;=\; \gamma/\nu~.
\label{hyps}
\ee
As a consequence only two critical exponents are independent. Because of
the hyperscaling relations and the already implicitly assumed equality
of the critical exponents above and below $T_c$ one can construct a
multitude of universal amplitude ratios \cite{Privman:1991} (see also the 
discussion in Ref.\ \cite{Pelissetto:2000ek}). The following list of
ratios contains those which we will determine here
\ba
&\!\!\!R_{\chi} \; =\; C^+ D_c B^{\delta-1}~,\, \quad 
&U_{\xi}^c \; \;= \; \xi^c_T /\xi^c_L ~,
\label{rur}\\
&Q_c\; =\; B^2 (\xi^+)^d /C^+ ~, \quad
&Q_2^{L,T} \; =\; (\xi^c_{L,T} /\xi^+)^{\gamma/\nu} C^+/ C^c~.
\label{qratios}
\ea

The critical behaviour of the magnetization in the vicinity of $T_c$
is more generally described by the magnetic equation of state. In its 
Widom-Griffiths form it is given by \cite{Widom:1965,Griffiths:1967}
\be
y \; =\; f(x)~,
\label{wigr}
\ee
where
\be
y \equiv h/M^{\delta}~, \quad x \equiv \bar t/M^{1/\beta}~.
\label{xandy}
\ee
The variables $\bar t$ and $h$ are the normalized reduced temperature
$\bar t= tT_c/T_0$ and magnetic field $h=H/H_0$, which are chosen such
as to fulfill the standard normalization conditions
\be
f(0)\; =\; 1~, \quad f(-1)\; =\; 0~,
\label{normf}
\ee
which imply
\be
M(t=0) = h^{1/\delta} \quad {\rm and } \quad H_0 = D_c~,
\label{normh}
\ee
\be
M(h=0) = (-\bar t\;)^{\beta} \quad {\rm and } \quad T_0 = B^{-1/\beta}T_c~.
\label{normt}
\ee
Possible dependencies on irrelevant scaling fields
and exponents are however not taken into account in Eq.\ (\ref{wigr}), 
the function $f(x)$ is universal. Another way to express the 
dependence of the magnetization on $\bar t$ and $h$ is
\be
M\;=\;h^{1/\delta} f_G(\bar t/h^{1/\beta\delta})~,
\label{mscale}
\ee
where $f_G$ is again a universal scaling function. The two forms 
(\ref{wigr}) and (\ref{mscale}) are of course equivalent. The
function $f_G(z)$ and its argument $z$ are related to $x$ and $y$ by
\be
f_G = y^{-1/\delta}~, \quad z \equiv \bar t/h^{1/\beta\delta}
= xy^{-1/\beta\delta}~.
\label{xyzf}
\ee
Correspondingly the normalization conditions (\ref{normf}) translate into
\be
f_G(0)\; =\; 1~,\quad {\rm and}\quad f_G(z) {\raisebox{-1ex}{$
\stackrel{\displaystyle\longrightarrow}{\scriptstyle z \rightarrow -\infty}$}}
(-z)^{\beta}~.
\label{normfg}
\ee
Since the susceptibility $\chi_L$ is the derivative of $M$ with respect to
$H$ we obtain from Eq.\ (\ref{mscale})
\be
\chi_L = {\partial M \over \partial H} = {h^{1/\delta -1} \over H_0}
 f_{\chi}(z)~,
\label{cscale}
\ee
with 
\be
f_{\chi}(z) = {1 \over \delta} \left( f_G(z) - {z \over \beta} f_G\p (z)
\right)~.
\label{fchi}
\ee
For $H\to 0$ at fixed $t>0$, that is for $z\rightarrow \infty$,
the leading asymptotic term of $f_{\chi}$ is determined by Eq.\ (\ref{chicr})
\be
f_{\chi} (z)\; {\raisebox{-1ex}{$\stackrel 
{\displaystyle =}{\scriptstyle z \rightarrow \infty}$}}
\;  C^+ D_c B^{\delta-1} z^{-\gamma}\;=\;R_{\chi} z^{-\gamma} ~.
\label{fcasy}
\ee
For $z\rightarrow \infty$
the leading terms of $f_G$ and $f_{\chi}$ are identical, because
for $T>T_c$ and small magnetic field $M$ is proportional to $H$.

Like for $M$ and $\chi_L$ the dependence 
of the correlation lengths on $\bar t$ and $h$ in the critical region
and the thermodynamic limit is given in terms of scaling functions
$g_{\xi}^{L,T} (z)$ by
\be 
\xi_{L,T}  \; =\; h^{-\nu_c} g_{\xi}^{L,T} (z)~.
\label{xiscale}
\ee
These functions are universal except for a normalization
factor. On the critical line $\bar t=0$ or $z=0$ we find from (\ref{xicrh})
\be
g_{\xi}^{L,T} (0) \; =\; \xi^c_{L,T} D_c^{-\nu_c}
\; =\;\xi^c_{L,T} (B^c)^{\nu/\beta}~,
\label{gxi0}
\ee
and from (\ref{xicr}) the asymptotic behaviour at $z \to  \infty$
(which is the same for both correlation lengths),
\be
g_{\xi}^{L,T} (z) \; {\raisebox{-1ex}{$\stackrel 
{\displaystyle =}{\scriptstyle z \rightarrow \infty}$}} \;
\xi^+ B^{\nu/\beta} z^{-\nu}~.
\label{gxiasy}
\ee
Indeed, the ratios of the amplitude for $z\to \infty$ in (\ref{gxiasy}) 
and the $g_{\xi}^{L,T} (0)$ are universal
\be
{\xi^{+} B^{\nu/\beta} \over \xi^c_{L,T} (B^c)^{\nu/\beta}} \; =\; 
\left({ \delta R_{\chi} \over Q_2^{L,T}} \right)^{\nu/\gamma} ~, 
\label{univers}
\ee
whereas the $g_{\xi}^{L,T} (0)$ itself are not. In analogy to the 
Ising case discussed in Ref.\ \cite{Engels:2002fi} we therefore define
two universal scaling functions by
\be
{\hat g}_{\xi}^{L,T}(z) \; = \; g_{\xi}^{L,T}(z)/g_{\xi}^{L,T}(0)~.
\label{gnorm}
\ee


\section{Numerical Details}
\label{section:update}

\n All our simulations were done on three-dimensional lattices with periodic
boundary conditions and linear extensions $L=48,72,96$ and 120. As in
Ref.\ \cite{Engels:1999wf} we have used the Wolff single cluster algorithm
\cite{Wolff:1988uh}, which was modified to include a non-zero magnetic field
\cite{Dimitrovic:yd}. In order to reduce the integrated 
autocorrelation time $\tau_{int}$ we 
performed 100-3000 cluster updates between two measurements for $H>0$, such
that we obtained
$\tau_{int}\ltapprox 3$ for $L= 48$ and $\tau_{int}\ltapprox 5$ for
$L\ge 72$. For $H=0, T>T_c$ and large lattices the number of cluster updates 
was increased up to 10000. In general we made 20000 measurements for each
fixed $H$ and $J$. We use the value for $J_c=T_c^{-1}=0.93590$ from Ref.\
\cite{Oevers:1996dt}. The coupling constant region which we have explored
was $0.9 \le J \le 1.2$, the magnetic field was varied from $H=0$ to 
$H=0.01$.  

\subsection{Measurement of the Correlation Lengths}
\label{section:tryxi}

Instead of using correlation functions of the individual spins it is more
favourable to consider spin averages over planes and their respective
correlation functions. For example, the spin average over the $(x,y)$-plane
at position $z$ is defined by
\be
{\vec S}_z =  { 1 \over L^2 } \sum_{x,y} \phv_{\vec x}~.
\label{spinav}
\ee
The spin averages have again longitudinal and transverse components
\be
{\vec S}_z \; =\; S_z^{\parallel} \vec e_H +
{\vec S}_z^{\perp} ~,
\label{Sdecomp}
\ee
The expectation values of the components are independent
of $z$ and equal to those of the respective lattice averages 
\be
\langle S_z^{\parallel} \rangle =\langle \phi^{\parallel} \rangle =M~,
\quad {\rm and} \quad \langle  {\vec S}_z^{\perp}\rangle =
 \langle\phv^{\perp}\rangle =0~.
\label{sexp}
\ee 
Correspondingly, we define now plane-correlation functions 
${\bar G}_{L,T}(z)$ by
\ba
{\bar G}_L(z)\!\!& =&\!\! L^2 \left (\langle S_0^{\parallel}
S_z^{\parallel} \rangle - M^2 \right)\label{planel}\\
{\bar G}_T(z)\!\!& =&\!\! {L^2 \over 3}
\langle {\vec S}_0^{\perp} {\vec S}_z^{\perp}\rangle ~.
\label{planet}
\ea
Here, $z$ is the distance between the two planes. Instead of choosing 
the $z$-direction as normal to the plane one can as well take the 
$x$- or $y$-directions. Accordingly, we enhance the accuracy of the 
correlation function data by averaging over all three directions 
and all possible translations. The correlators are symmetric
and periodic functions of the distance $\tau$ between the planes,
the factor $L^2$ on the right-hand sides of (\ref{planel})
and (\ref{planet}) ensures the relation
\be
\chi_{L,T} = \sum_{\tau=0}^{L-1} {\bar G}_{L,T}(\tau)~.
\label{fluc}
\ee
Like the point-correlation functions in Eq.\ (\ref{xiexp}) the 
plane correlators ${\bar G}_{L,T}(\tau)$ decay exponentially. In order 
to obtain the corresponding exponential correlation length $\xi$ from 
${\bar G}(\tau)$ we therefore make the ansatz
\be
{\bar G}(\tau ) = A \left[\: \exp (-\tau /\xi ) + 
\exp (-(L-\tau )/\xi )\: \right]~,
\label{expans}
\ee
and then try to fit the data for the correlation functions in an
appropriate $\tau$-range. The ansatz (\ref{expans}) implies of course,
that there are no additional excitations contributing to 
${\bar G}(\tau)$. Inspired by the experiences reported in the Ising case
\cite{Caselle:1997hs,Engels:2002fi}, we proceed in the following way.
First we calculate an effective correlation length $\xi^{eff}(\tau)$ from 
(\ref{expans}) using only the correlators at $\tau$ and $\tau +1$. For
$\tau \ll L$ this correlation length is approximately given by
\be
\xi^{eff}(\tau) = {-1 \over \ln ({\bar G}(\tau +1)/{\bar G}(\tau)) }~.
\label{xieff}
\ee
\begin{figure}[t!]
\begin{center}
   \epsfig{bbllx=63,bblly=280,bburx=516,bbury=563,
       file=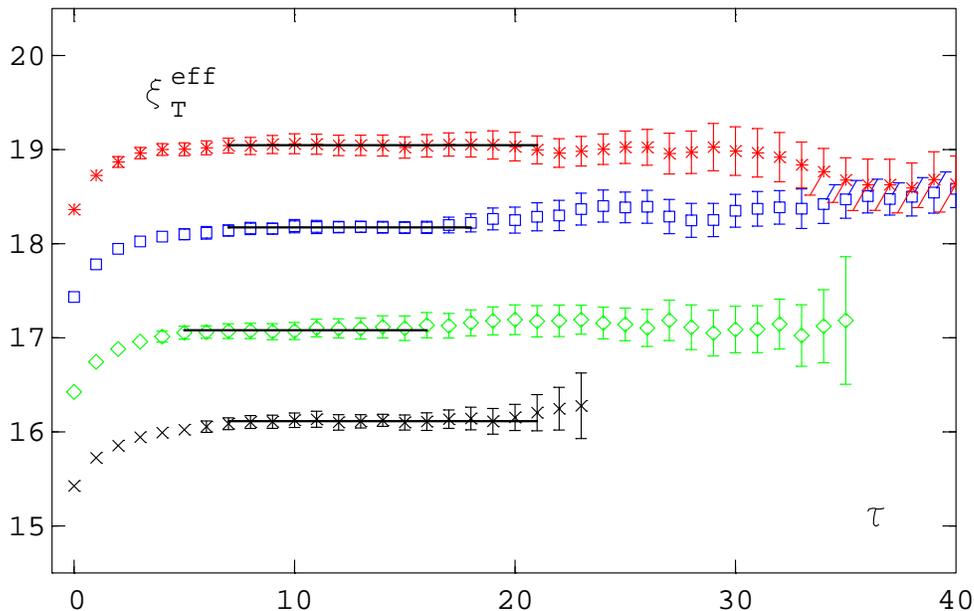, width=120mm}
\end{center}
\caption{The effective transverse correlation length $\xi^{eff}_T$ 
at $J=0.95$ and $H=0.001$ for $L= 48$ (crosses), 72 (diamonds),
96 (squares) and 120 (stars) as a function of $\tau$. 
The lines show the corresponding $\tau$-ranges where fits to Eq.\  
(\ref{expans}) were performed, their heights the resulting fit values.
The data for $L=72,96$ and 120 have been shifted upwards by 1,\,2 and 3
for better visibility.}
\label{fig:plateau}
\end{figure}
\n As an example, we show
in Fig.\ \ref{fig:plateau} a typical result for $\xi^{eff}(\tau)$
at $J=0.95,\, H=0.001$, that is in the low tempera\-ture region, from the 
transverse correlation functions on lattices with $L=48,72,96$ and 120. 
With increasing $\tau$ also $\xi^{eff}$ increases and eventually reaches 
a plateau inside its error bars. Most likely, the lower values of 
$\xi^{eff}$ at small $\tau$ are due to higher excitations. At large 
distances $\tau$ the resulting $\xi^{eff}$ start to fluctuate when the
relative errors of the data become too large. The two limits define an
intermediate $\tau$-range where a global fit with the ansatz (\ref{expans})  
can be used to estimate the exponential correlation length $\xi$.
It is clear, that due to the strong correlations between the values 
${\bar G}(\tau)$ for different $\tau$, the result from a fit in the whole
range of the plateau must be essentially the same as that of one 
(low) $\tau$-value in the plateau. That is indeed the case:
the two estimates for $\xi$ agree within the error bars and the errors
themselves are of about the same size, or somewhat larger for the plateau
fits. 
\begin{table}[t]
\begin{center}
\begin{tabular}{|l||c|l|c||c|c|c|}
\hline
$J$ & $\xi_T^{eff}(\tau_a)$ & ~~~~~$\xi_T$ &$[\tau_a,\tau_b]$
 & $\xi_L^{eff}(\tau_a)$ & $\xi_L$ & $[\tau_a,\tau_b]$ \\
\hline
 0.95   & 16.054(69) & 16.080(105) & 5-16 & - & - & - \\
 0.9359 & 13.447(46) & 13.449(49)  & 6-21 & 6.738(55) & 6.767(64) & 8-17 \\
 0.93   & 11.789(36) & 11.813(32)  & 6-26 & 7.594(57) & 7.647(60) & 8-19 \\
 0.92   & ~8.563(32) & ~8.592(26)  & 3-17 & 7.705(54) & 7.720(60) & 4-15 \\
\hline
\end{tabular}
\end{center}
\vspace{-0.2cm}
\caption{Correlation length estimates for different $J$ and $H=0.001$:
$\xi^{eff}(\tau_a)$ is the value at the lower end of the plateau
region $[\tau_a,\tau_b]$, $\xi$ the value from fits to the ansatz 
(\ref{expans}) in the whole plateau region.}
\label{tab:xico}
\end{table}
 Using the latter makes the result less dependent on the special
choice of $\tau$, where $\xi^{eff}$ is evaluated, in case the 
$\xi^{eff}$-values are slightly fluctuating. In order to give an impression
of the different $\xi$-estimates, their errors and the respective
$\tau$-ranges we show in Table \ref{tab:xico} some representative results 
for $T$ below, at and above $T_c$ and $H=0.001$ on a lattice with 
$L=72$ (at $J=0.92$ with $L=48$). \\ \\
\indent With our method we found reliable values for the 
exponential transverse correlation length $\xi_T$ at all $T$.
In the case of the longitudinal correlation length $\xi_L$ the same is
true in the high temperature phase $T>T_c$. However, at $T_c$ and 
slightly below the critical point at $H>0$ there are already indications
for the presence of higher states. Deeper in the low temperature phase it
becomes impossible to estimate the longitudinal correlation length:
no plateau appears, there is a dense spectrum of states contributing to
the longitudinal correlator.   

\section{Critical Amplitudes and Universal Ratios}
\label{section:amandrat}
\n In Ref.\ \cite{Engels:1999wf} the spontaneous magnetization
$M(T,0)$ was calculated at several temperatures on the coexistence line,
taking advantage of the Goldstone effect on $M$. From these results the
critical amplitude $B$ of Eq.\ (\ref{mcr}) and the normalization $T_0$
could be determined to $B=0.9916(5)$ and $T_0=1.093(2)$ using $\beta=0.380$.
We use these values in the following. 
In the same paper the amplitude $B^c$ and the critical exponent $\delta$
were determined from fits to the critical behaviour of $M$ on the critical 
line. Since we have done new simulations on the critical line with higher
statistics and at more $H$-values on the larger lattices than in
\cite{Engels:1999wf} we could improve the analysis with the new
magnetization data.  
\newpage

\subsection{Results from the Critical Line}
\label{section:critline}
\begin{figure}[t!]
\begin{center}
   \epsfig{bbllx=63,bblly=280,bburx=516,bbury=563,
       file=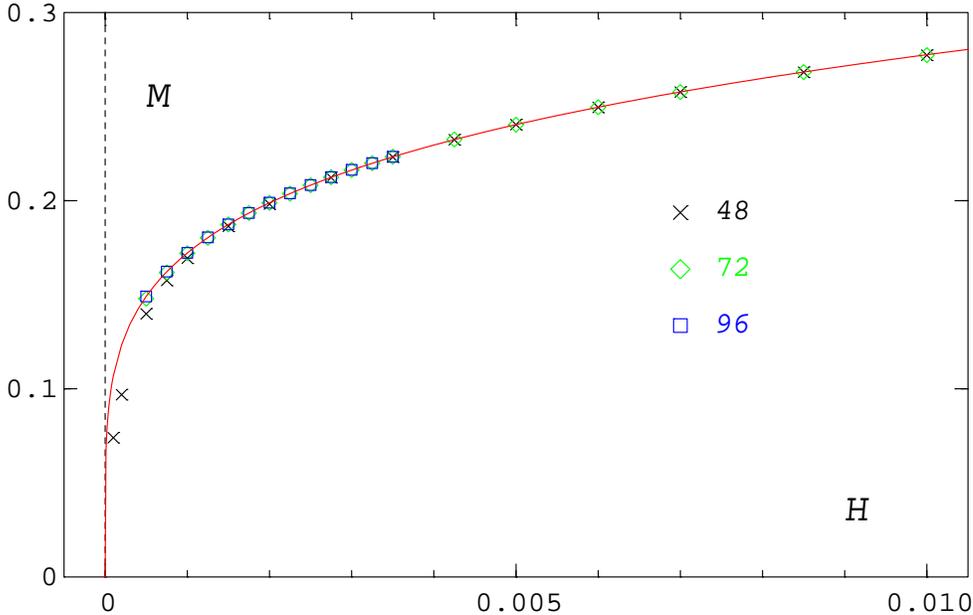, width=120mm}
\end{center}
\caption{The magnetization $M$ at $T_c$ as a function of the magnetic field
$H$ for $L=48,72$ and 96. The solid line shows the fit (\ref{mhcorr}).}
\label{fig:mtzero}
\end{figure}
\begin{figure}[t!]
\begin{center}
   \epsfig{bbllx=63,bblly=280,bburx=516,bbury=563,
       file=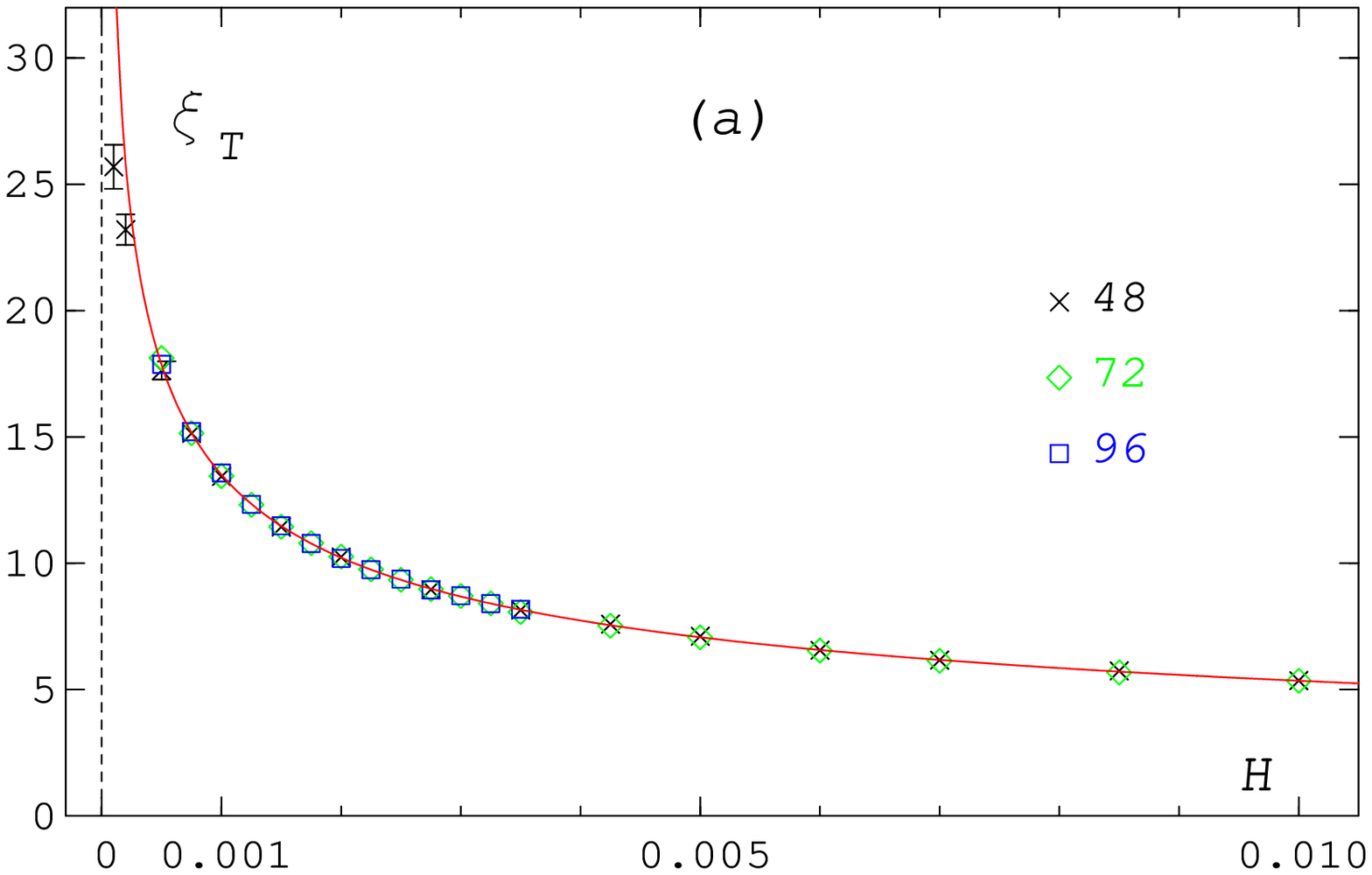, width=120mm}
\end{center}
\vspace{0.8mm}
\begin{center}
   \epsfig{bbllx=63,bblly=280,bburx=516,bbury=563,
       file=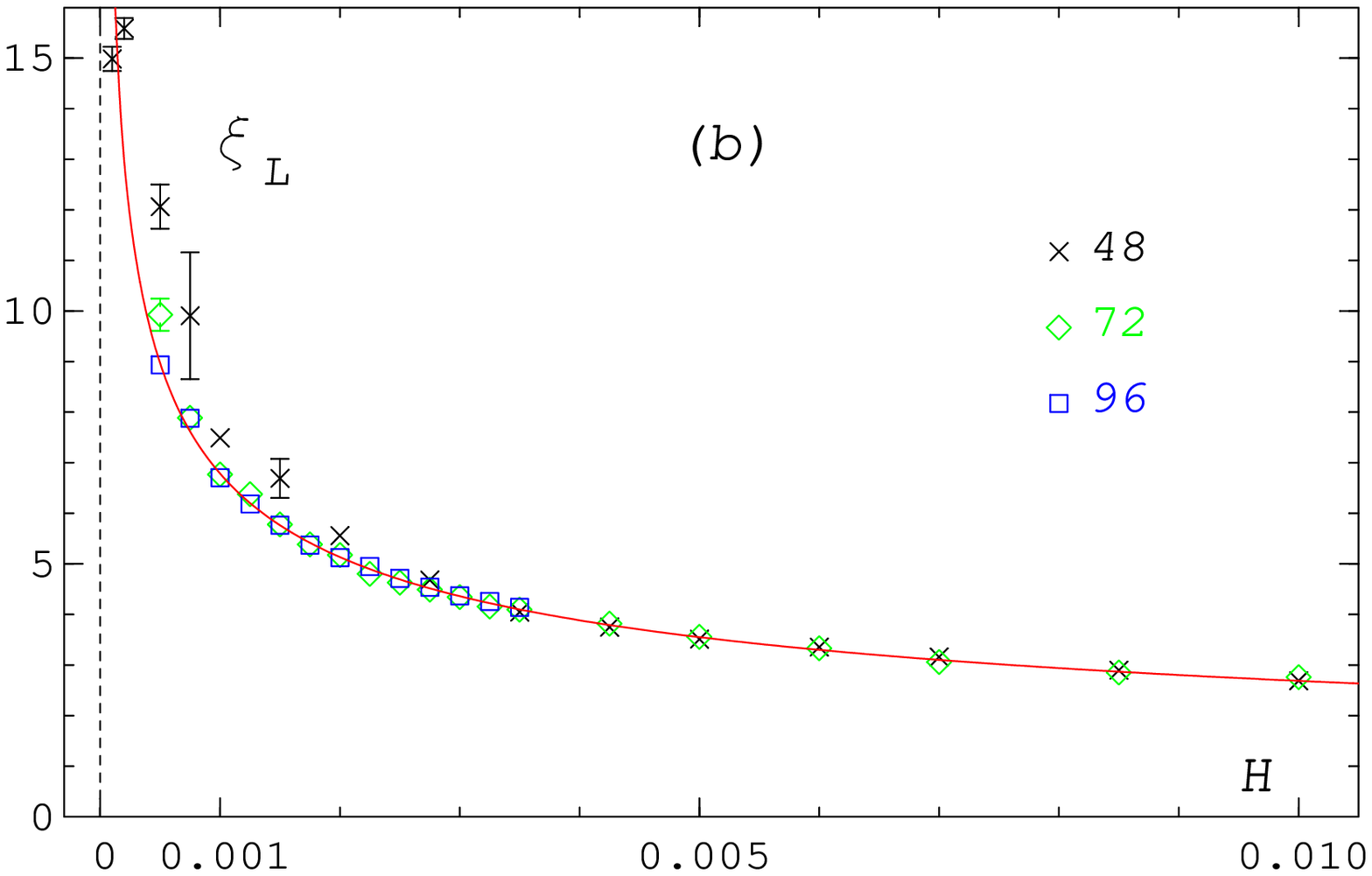, width=120mm}
\end{center}
\caption{The correlation lengths $\xi_T$ (a) and $\xi_L$ (b) at $T_c$ 
for $L= 48,72$ and 96 as a function of $H$. The lines are the fits  
(\ref{fitcor}) with $\xi_{T,L}^c$ from (\ref{xicamp}) and $c_{T,L}=0$.}
\label{fig:cole}
\end{figure}

\indent On the critical isotherm, that is at $T_c$, we have measured the 
magnetization for $H>0$ on lattices with $L=48,72$ and 96. The results are
shown in Fig.\ \ref{fig:mtzero}. We observe essentially only for $L=48$
a noticeable finite size dependence close to $H=0$. Like in 
\cite{Engels:1999wf} we have made various fits to the data from the
largest lattices in the $H$-interval $[0.00075,\;0.00425]$ with the 
simple ansatz
\be
M\; =\; B^c H^{1/\delta}~.
\label{mhcorr}
\ee
The resulting values for the amplitude and the exponent are (with 
$\chi^2/N_f\approx 1.7$)
\be 
B^c \; = \; 0.721(2)\quad {\rm and }\quad \delta \;=\;4.824(9)~,
\label{bece}
\ee   
and lead to $D_c= H_0 =4.845(66)$. The solid curve in Fig.\
\ref{fig:mtzero} shows this fit. It also represents the data at higher
magnetic fields very well. From these additional data a marginal negative
correction-to-scaling term may be inferred, which is however irrelevant
at low $H$. Compared to the corresponding fit in \cite{Engels:1999wf},
where $\delta \;=\;4.86(1)$, we note that our new result for
$\delta$ is closer to the value 4.789(6) from Ref.\
\cite{Hasenbusch:2000ph}. The latter value was however deduced from
calculations at zero magnetic field. 
All the other critical exponents which were needed in our calculations 
have been derived from the hyperscaling relations (\ref{hyps}) and
$\beta=0.380,\: \delta=4.824\:$. We use therefore
\be
\nu= 0.7377~,\quad \gamma=1.4531~, \quad \nu_c=0.4024~.
\label{expo}
\ee
\n The data for the transverse and longitudinal correlation lengths on the
critical line are shown in Fig.\ \ref{fig:cole}$\;$(a) and (b). For the
transverse correlation length $\xi_T$ we find only finite size effects
very close to $H=0$.
The longitudinal correlation length $\xi_L$ exhibits
fluctuations and a systematic deviation to higher $\xi_L$-values, when the
magnetic field is decreasing. The smaller the lattice, the earlier this
behaviour sets in. In order to determine the amplitudes we have fitted our 
results to the following form with $\omega=0.8$
\be
\xi_{T,L} \; = \;\xi_{T,L}^c H^{-\nu_c} \left( 1 + c_{T,L} H^{\omega\nu_c}
\right)~.  
\label{fitcor}
\ee 
In both cases we have taken various subsets of the data for the fits in 
the $H$-interval [0.0005,0.005]. It turned out that the correction
terms are zero within their error bars and that fits with  $c_{T,L}\equiv 0$
work just as well. The $\chi^2/N_f$ varies between 0.5 and 0.8\,. We find for
the amplitudes
\be
\xi_T^c \; =\;0.838(1)~, \quad \xi_L^c \; =\; 0.421(2)~.
\label{xicamp}
\ee 
As can be seen from Fig.\ \ref{fig:cole}$\;$(a) and (b), the corresponding
fits describe the data also in the higher $H$-range up to $0.01\:$. 
The ratio of the two correlation lengths is remarkably close to 2
\be
U_{\xi}^c \; =\;\xi_T^c/\xi_L^c \; =\; 1.99(1)~,
\label{ratxi}
\ee
that is, the expectation for the ratio from Eq.\ (\ref{corrcon}) close to 
the coexistence line is fulfilled also at $T_c$. A similar observation has 
been made for the three-dimensional $O(2)$ model \cite{Cucchieri:2002hu}.

\subsection{Results for $T>T_c$ and $H=0$}
\label{section:critiso}

\begin{figure}[t]
\begin{center}
   \epsfig{bbllx=63,bblly=280,bburx=516,bbury=563,
       file=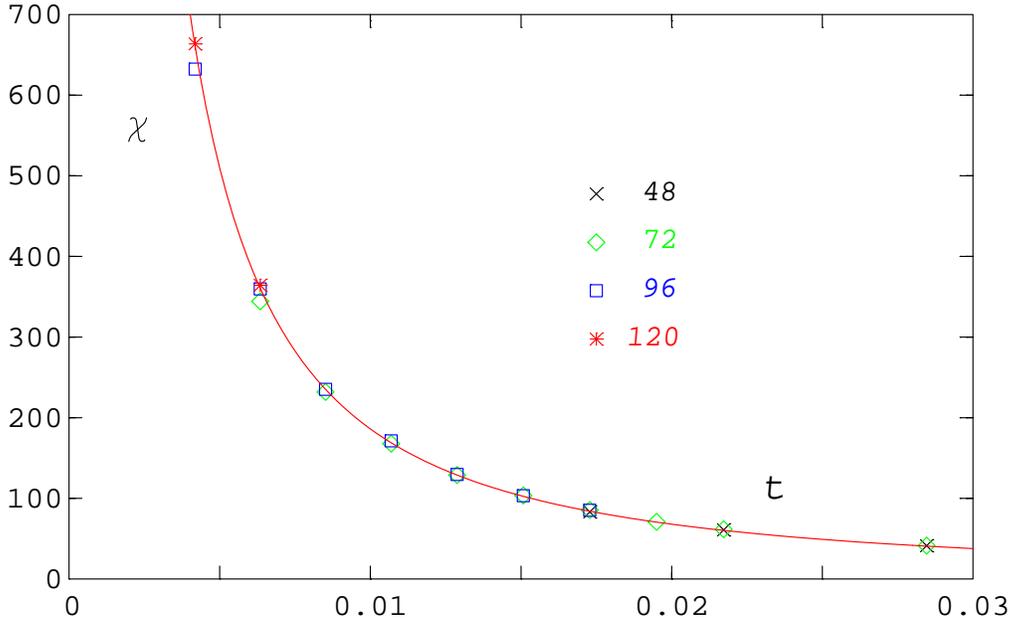, width=120mm}
\end{center}
\caption{The susceptibility above $T_c$ at $H=0$ for $L=48,72,96$ and 120
versus the reduced temperature $t$. The solid line shows the leading term
of the fit to Eq.\ (\ref{bchim}).}
\label{fig:chihzero}
\end{figure}
Our next aim is the determination of the critical amplitudes $C^+$ and
$\xi^+$. To this end we have evaluated the data of the susceptibility and
of the correlation length above the critical temperature for
$H=0$. The data points for $\chi=\chi_L=\chi_T$ are plotted in Fig.\ 
\ref{fig:chihzero} as a function of $t$. We have made various fits with
the ansatz
\be
\chi\; =\; C^+ t^{-\gamma} [1 +C_1^+ t^{\omega\nu}]~,
\label{bchim}
\ee
in the $J$-interval [0.91,0.932], which corresponds to 
$0.00415\le t \le 0.0285$. All fits, including a correction term
$\sim C_1^+$ or not, are compatible with the critical amplitude result
\be
C^+ \: = \: 0.231(2)~.
\label{cep}
\ee
\begin{figure}[t!]
\begin{center}
   \epsfig{bbllx=63,bblly=280,bburx=516,bbury=563,
       file=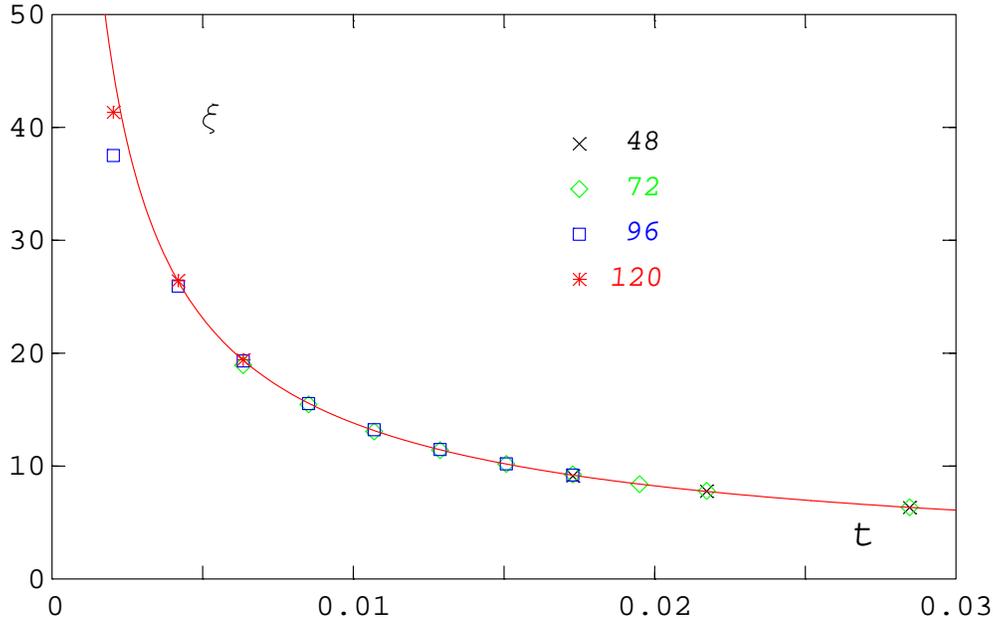, width=120mm}
\end{center}
\caption{The correlation length $\xi$ above $T_c$ at $H=0$ for
$L=48,72,96$ and 120 versus $t$. The solid line shows the fit from
Eq.\ (\ref{xipfit}).}
\label{fig:xiat}
\end{figure}
\noindent
For the correction amplitude we find $C_1^+=0.14(7)$, and $\chi^2/N_f$ is in
the range $[1.3,1.8]$\,. As can be seen in Fig.\ \ref{fig:chihzero}, where 
only the leading term is plotted, the correction contribution is marginal. 

The correlation length data for $H=0$ above the critical temperature
are shown in Fig.\ \ref{fig:xiat}. Like for the susceptibility we
have made an ansatz including a correction-to-scaling term
\be
\xi = \xi^+ t^{-\nu} [ 1 + \xi^+_1 t^{\omega\nu}]~.
\label{xiant}
\ee  
\vspace*{1mm}
\noindent Different fits in the same $J$-interval [0.91,\,0.932], which
was already used in the case of the susceptibility, lead to the result
\be
\xi = 0.466(2) t^{-\nu} [ 1 - 0.12(4) t^{\omega\nu}]~,
\label{xipfit}
\ee
with a $\chi^2/N_f$ between 0.8 and 1.2\,.
The critical amplitude of the correlation length is therefore
\be
\xi^+ = 0.466(2)~.
\label{xipamp}
\ee

\subsection{Universal Amplitude Ratios}
\label{section:ratios}

\n The critical amplitudes $B^c,\:\xi^c_{L,T},\:C^+,\:\xi^+$, 
which we have just determined, and $B$ from \cite{Engels:1999wf} enable
us to fix the necessary normalizations of the scaling functions and
in addition we can calculate the universal amplitude ratios defined in
Eqs.\ (\ref{rur}) and (\ref{qratios}). The ratio $U_{\xi}^c$ has already
been discussed. We obtain for the other ratios
\be
R_{\chi} = 1.084(18)~,\quad Q_c = 0.431(9)~,\quad Q_2^T =4.91(8)~,
\quad Q_2^L= 1.265(24)~.
\label{Rnum}
\ee   
For the ratio $R_{\chi}$ several alternative values exist: from the 
parametrization of the equation of state in \cite{Engels:1999wf} one
gets 1.126(9), from another parametrization in \cite{Toldin:2003hq}
1.12(11), from the $1/N$-expansion of Oku and Okabe \cite{Oku:1979ht}
1.098, and from the $\epsilon$-expansion 1.239 \cite{Privman:1991}.
Our finding for $Q_c$ is well in accord with the result 0.44(2)
of \cite{Toldin:2003hq}. We could not find competing values for the 
ratios $Q_2^{L,T}$ in the literature.


\section{The Scaling Functions}
\label{section:scalef}

\n In Ref.\ \cite{Engels:1999wf} the equation of state in the 
Widom-Griffiths form was parametrized 
by a combination of a small-$x$ (low temperature) form $x_s(y)$,
which was inspired by the approximation of Wallace and Zia 
\cite{Wallace:1975} close to the coexistence line ($x=-1;~y=0$)
\be
x_s(y)+1 \;=\; ({\widetilde c_1} \,+\, {\widetilde d_3})\,y \,+\,
             {\widetilde c_2}\,y^{1/2} \,+\, 
             {\widetilde d_2}\,y^{3/2} \;,
\label{PTform}
\ee
and a large-$x$ (high temperature) form $x_l(y)$ derived from  
Griffiths's analyticity condition \cite{Griffiths:1967}
\be
x_l(y)\;=\; a\, y^{1/\gamma} \,+\, b\,y^{(1-2\beta)/\gamma}~.
\label{highx}
\ee
The two parts are interpolated smoothly by the ansatz
\be
x(y) \;=\; x_s(y)\,\frac{y_0^p}{y_0^p + y^p} \,+\,
           x_l(y)\,\frac{y^p}{y_0^p + y^p}~,
\label{totalfit}
\ee
from which the total scaling function is obtained. Since we have
changed the values of $H_0$ and of $\delta$ in comparison to Ref.\
\cite{Engels:1999wf}, we have redone the corresponding 
parametrization fits. With the new data we find
\be
{\widetilde c_1} \,+\, {\widetilde d_3}\;=\;0.359(10)~,\quad
 {\widetilde c_2}\;=\;0.666(6)~,
\label{newxs}
\ee
and ${\widetilde d_2}= 1-({\widetilde c_1}+{\widetilde d_3}+
{\widetilde c_2})$ is fixed by the normalization condition $y(0)=1$.
The new values for $a$ and $b$ are
\be
a\;=\;1.071(4)~, \quad b\;=\;-0.866(38)~.
\label{newxl}
\ee
The interpolation parameters $y_0=10.0$ and $p=3$ have been retained
unchanged. The modified values of $a$ and $\gamma$ result in a new 
estimate of $R_{\chi}$ 
\be
R_{\chi}\;=\;a^{\gamma}\;=\;1.105(4)~,
\label{rchin}
\ee
in nice agreement with the value we obtained in (\ref{Rnum}) directly
from the amplitudes.

\begin{figure}[tp]
\begin{center}
   \epsfig{bbllx=63,bblly=265,bburx=516,bbury=588,
       file=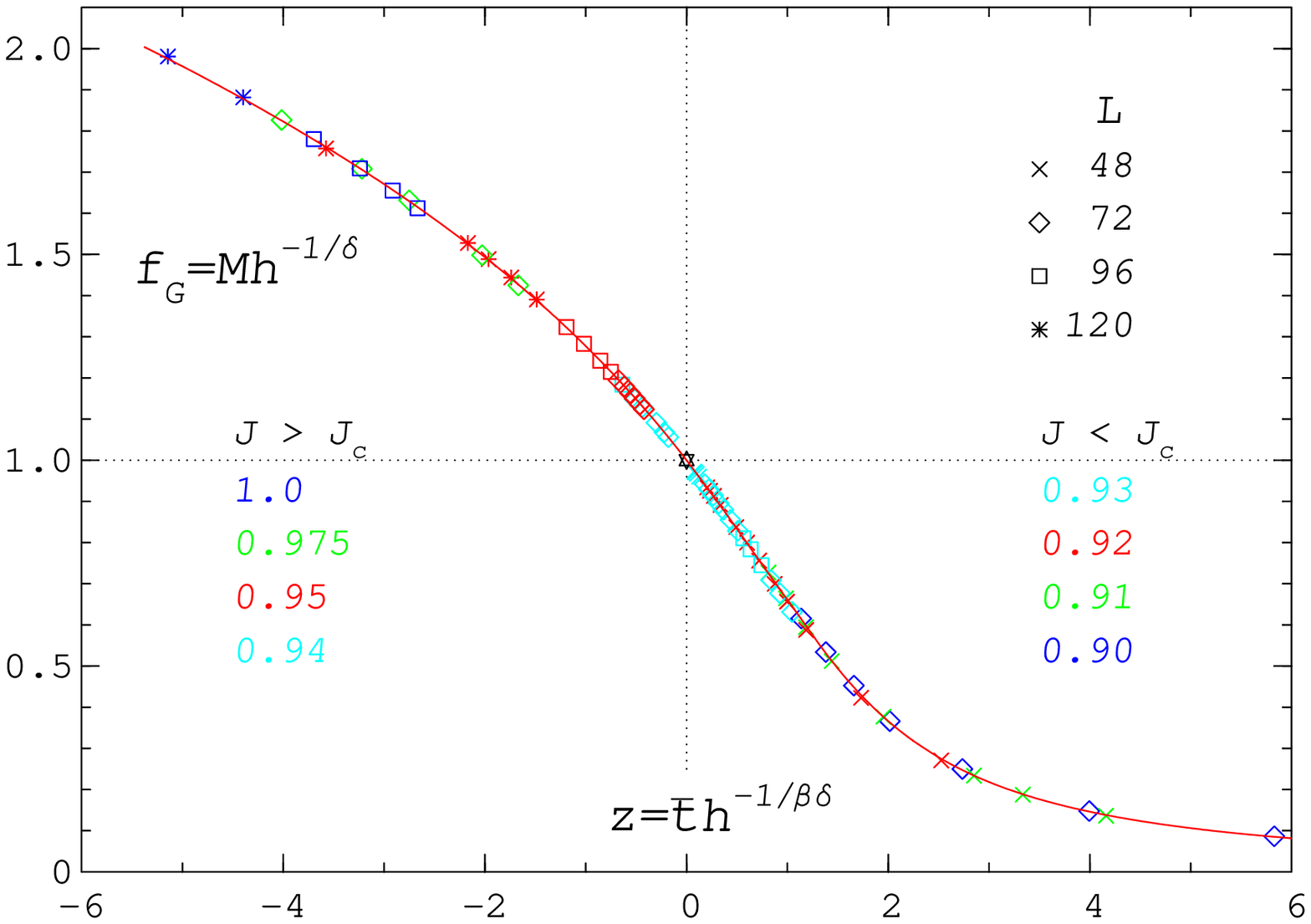, width=120mm}
\end{center}
\caption{The equation of state: $f_G=Mh^{-1/\delta}$ as a function of 
$z=\bar t h^{-1/\beta\delta}$. The solid line shows our parametrization.
The numbers refer to the different $J=1/T$-values of the data, the
dotted lines cross at the normalization point $f_G(0)=1$.}
\label{fig:fG(z)}
\begin{center}
   \epsfig{bbllx=63,bblly=265,bburx=516,bbury=588,
       file=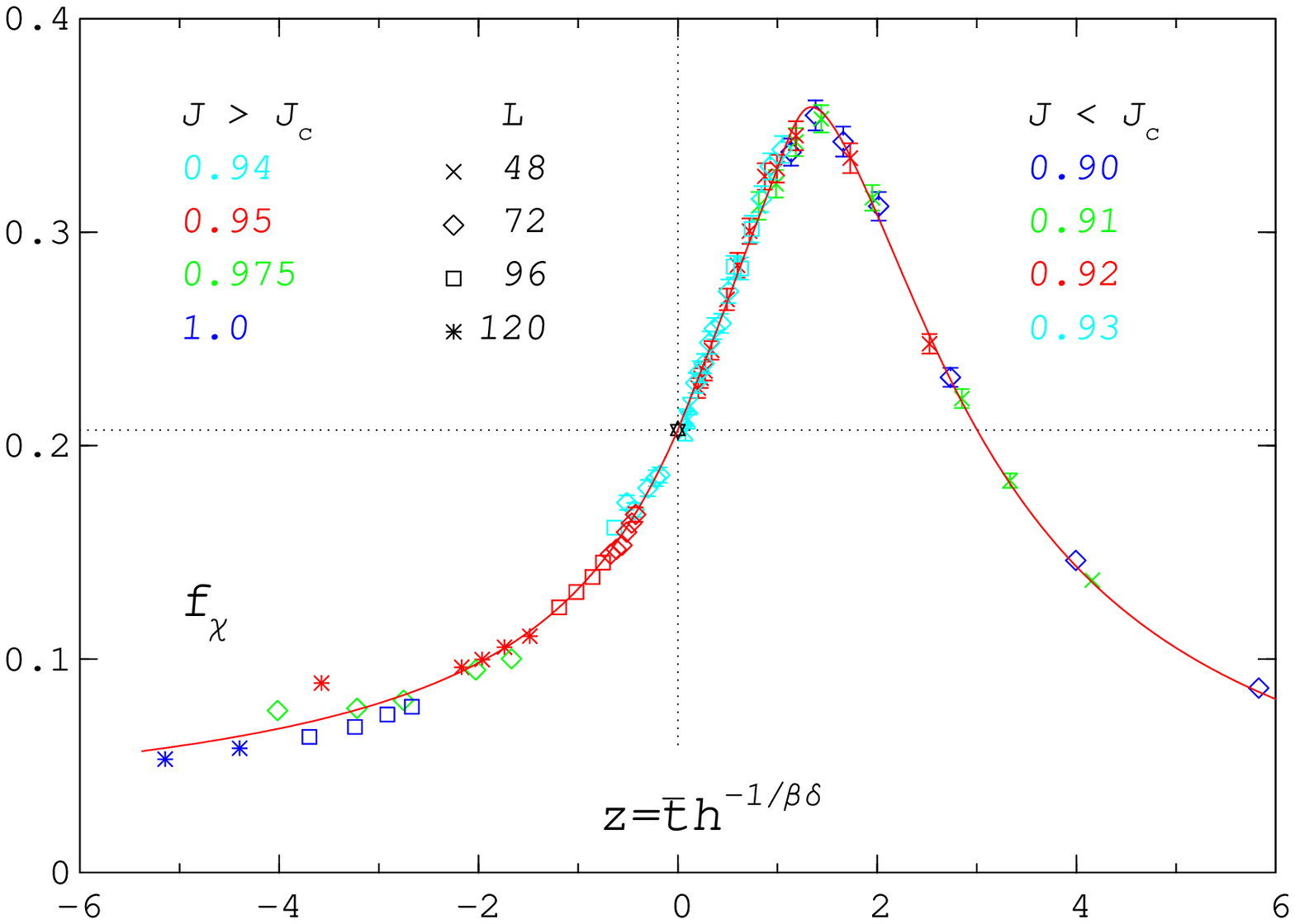, width=120mm}
\end{center}
\caption{The scaling function
$f_{\chi}(z)=\chi_L H_0 h^{1-1/\delta}$ versus $z=\bar th^{-1/\beta\delta}$.
The solid line is from our parametrization of the equation of state.
The numbers refer to the $J=1/T$-values of the data. The normalization is
$f_{\chi}(0)=1/\delta$.}
\label{fig:fchi}
\end{figure}
\indent In Fig.\ \ref{fig:fG(z)} we present the results for $f_G(z)$ obtained 
from the magnetization data in the coupling range  $0.90\le J \le 1.0\:$.
At first sight these data are scaling well. Since the data from the coupling 
$J=1.2$ showed already visible deviations from scaling behaviour we have
discarded them. Similarly, limitations were found for the scaling 
$H$-regions. The data for $f_G(z)$ were scaling in the low temperature
region only up to $H=0.005$ and in the high temperature region
for $J=0.93$ and  $J=0.92$ up to $H=0.05$, for $J=0.91$ and $J=0.90$
the scaling $H$-range extended only to $H=0.01$. Only data within these
parameter ranges are shown in Fig.\ \ref{fig:fG(z)}$\,.$ In the same 
figure we show the function $f_G(z)$, which one obtains with Eq.\ 
(\ref{xyzf}) from our parametrization for $x(y)$. It obviously describes
the data quite well.  
The scaling function $f_{\chi}(z)$ of the susceptibility is connected via
Eq.\ (\ref{fchi}) to the scaling function $f_G(z)$ of the magnetization. It
can therefore be calculated from the parametrization as well. On the other
hand we have direct data for $\chi_L$ from our simulations which allows for
another check of the scaling hypothesis and a comparison to the 
parametrization. In Fig.\ \ref{fig:fchi} we show the respective data for
the same $J$-values as in Fig.\ \ref{fig:fG(z)}. The data are not as 
accurate as those for the magnetization. We observe explicit scaling 
in the high temperature phase ($z>0$), but with decreasing temperature,
or larger $-z$, the data are spreading and no longer scaling. In particular
the data at the coupling $J=1.0$ seem to be already outside the critical
region. For $J=0.975$ and $J=0.95$ it is possible that at very small $H$
(large $-z$) finite size effects are responsible for that behaviour. 
The other data are coinciding in essence with our parametrization for
$z>-3$.  
\begin{figure}[t!]
\begin{center}
   \epsfig{bbllx=63,bblly=265,bburx=516,bbury=588,
       file=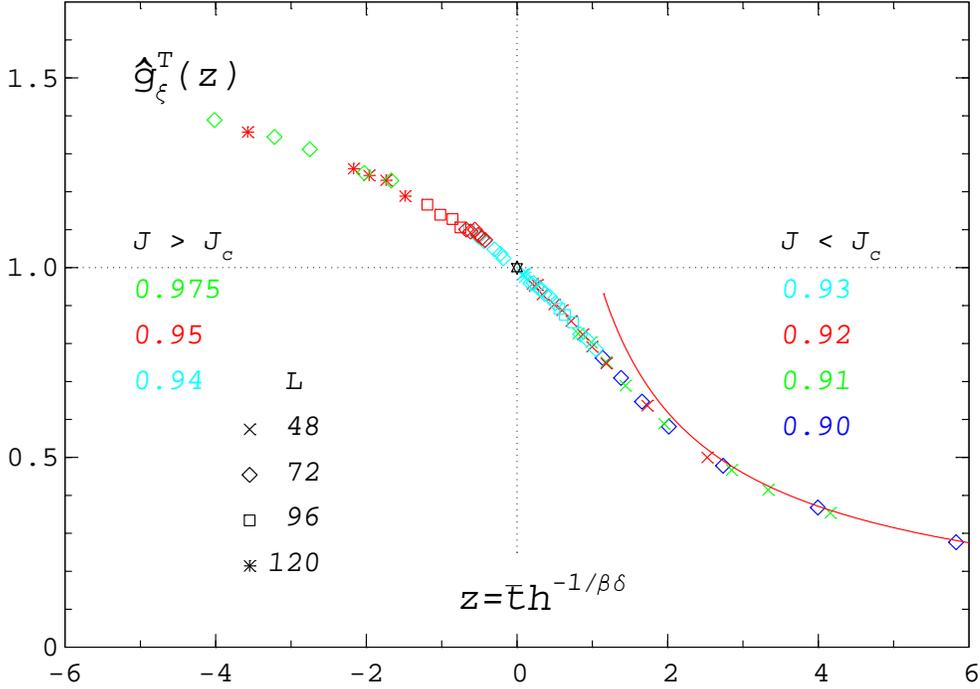, width=120mm}
\end{center}
\caption{The scaling function of the transverse correlation length
$\hat g_{\xi}^T(z)=\xi_T h^{\nu_c}/g_{\xi}^T(0)$. The solid line is the 
asymptotic form calculated from Eq.\ (\ref{gxiasy}). The numbers refer
to the different $J=1/T$-values of the data.}
\label{fig:gexit}
\end{figure}
\subsection{The Scaling Functions of the Correlation Lengths}
\label{section:xiscale}
\noindent In Fig.\ \ref{fig:gexit} we show the data for the normalized scaling 
function $\hat g_{\xi}^T(z)=\xi_T h^{\nu_c}/g_{\xi}^T(0)$. The determination of 
the transverse correlation length with the method described in Section
\ref{section:tryxi} was always possible. In each case a plateau in $\xi^{eff}$ 
could be found. The data shown in Fig.\ \ref{fig:gexit} correspond to the ones
presented in Fig.\ \ref{fig:fchi} for the scaling function of the
susceptibility, apart from the points for $J=1.0\,$, which were clearly outside
the scaling region. The normalization was calculated from Eq.\ (\ref{gxi0}) and
is $g_{\xi}^T(0)=0.444(3)$. As in the case of the magnetization and the
susceptibility we observe scaling of the correlation length data in a large
$z$-range. For $z>0$ we see an early approach to the asymptotic
form calculated from Eq.\ (\ref{gxiasy}). The shape of $\hat g_{\xi}^T(z)$
is very similar to that of $f_G(z)$. It is not by accident that we find this
behaviour but it is due to the relation (\ref{xichi}) between the transverse
correlation length and its susceptibility, which we discussed in Section 
\ref{section:Gold}$\:.$ We consider therefore the ratio of $\xi_T^2$ and 
$\chi_T$ and find
\be
{\xi_T^2 \over \chi_T} \;=\; (\xi^c_T)^2(B^c)^{-1} H^{-\eta\nu_c}
\cdot {(\hat g_{\xi}^T)^2 \over f_G}~,
\label{xisochi}
\ee
\begin{figure}[t!]
\begin{center}
   \epsfig{bbllx=63,bblly=265,bburx=516,bbury=588,
       file=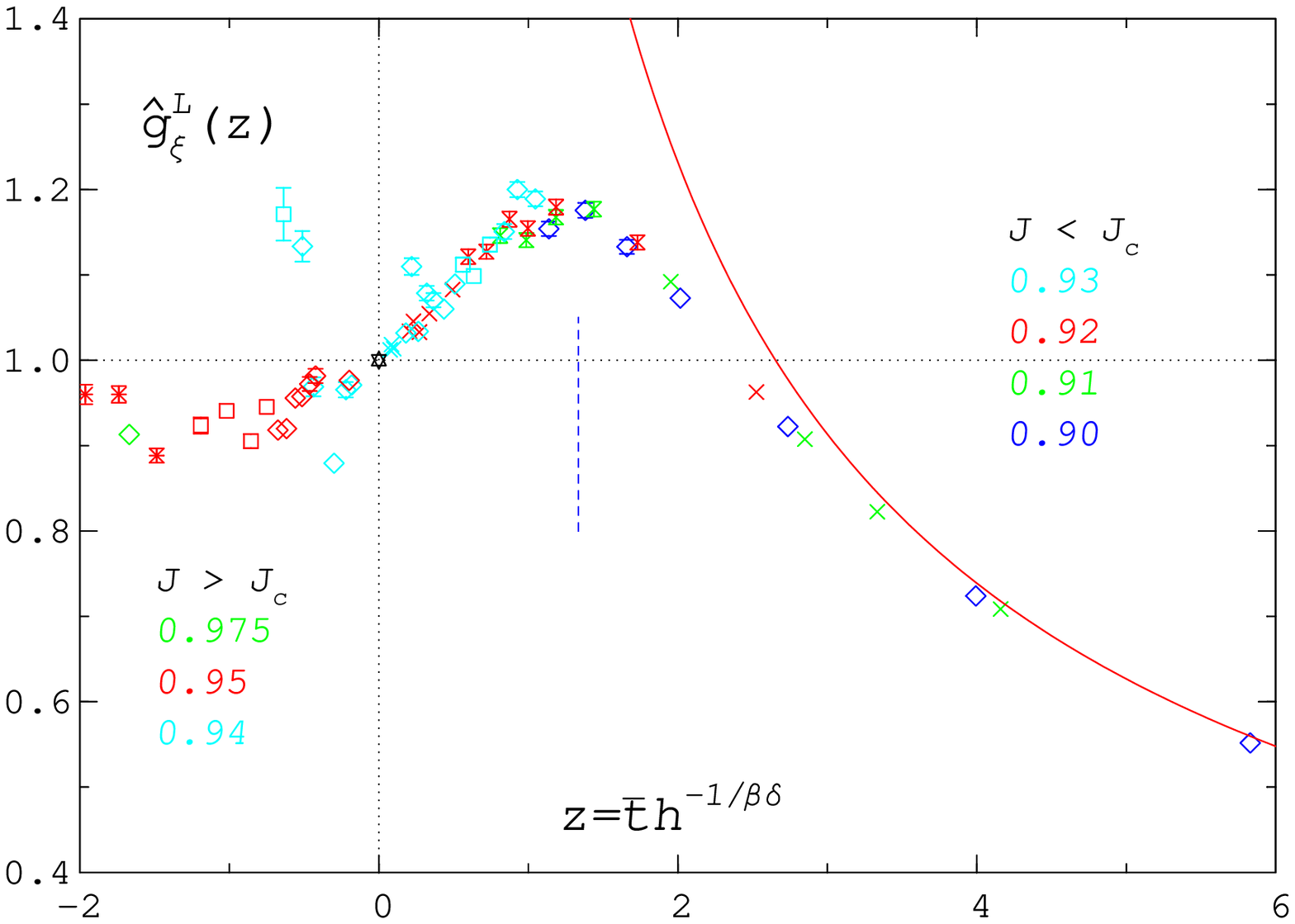, width=120mm}
\end{center}
\caption{The scaling function of the longitudinal correlation length
$\hat g_{\xi}^L(z)=\xi_L h^{\nu_c}/g_{\xi}^L(0)$. The solid line is 
obtained from the asymptotic form (\ref{gxiasy}). The numbers refer to
the different $J=1/T$-values. The dashed line indicates the peak 
position of $\chi_L$, the symbol notation is as in Fig.\ \ref{fig:gexit}.}
\label{fig:gexil}
\end{figure}
the same relation as in the Ising case \cite{Engels:2002fi}, execpt that
here $f_{\chi}$ is replaced by $f_G$, which plays the corresponding
part for the transverse correlation length. In the Ising case we
found the asymptotic behaviour
\be
{\hat g_{\xi}^2 \over f_{\chi}} \; {\raisebox{-1ex}{$\stackrel 
{\displaystyle \sim}{\scriptstyle z \rightarrow \pm\infty}$}} \;
 (\pm z)^{-\eta\nu}~.
\label{ratasy}
\ee
It is easy to show that $(\hat g_{\xi}^T)^2/ f_G$ for 
$z\rightarrow +\infty$ is proportional to $z^{-\eta\nu}$. If for 
$H\rightarrow 0$ the transverse correlation length behaves as
$\xi_T \sim H^{-1/2}$ - as expected because of the Goldstone effect -
then for $z\rightarrow -\infty$ the ratio $(\hat g_{\xi}^T)^2/ f_G$
must be proportional to $(-z)^{-\eta\nu}$, as in the Ising case. 
The condition for the asymptotic behaviour of the ratio in the broken 
phase and the $H^{-1/2}$-behaviour of $\xi_T$ are thus equivalent. 

\n Assuming the corresponding asymptotic behaviour for
$(\hat g_{\xi}^L)^2/ f_{\chi}$ would imply that $\xi_L \sim H^{-1/4}$ for
$T<T_c$, in contradiction to the expectation of a factor of 2 between the
two correlation lengths. The situation in the low temperature region is 
indeed difficult and as it seems we cannot clarify numerically the status
of $\xi_L$ there. In Fig.\ \ref{fig:gexil} we show our results for the
normalized scaling function $\hat g_{\xi}^L(z)=\xi_L h^{\nu_c}/g_{\xi}^L(0)$,
where $g_{\xi}^L(0)=0.223(2)$ was determined from Eq.\ (\ref{gxi0}). In the
high temperature region the longitudinal correlation length could be
calculated with sufficient accuracy, the finite size effects were negligible. 
Approaching the critical temperature ($z=0$) from above, the data become
more and more noisy and below $T_c$ we are essentially unable to determine  
$\xi_L$ and $\hat g_{\xi}^L(z)$: if one can find a plateau in $\xi^{eff}$
at all, then the results for different lattice sizes are often very different.
In addition scaling is completely lost already for small negative values of 
$z$. We have therefore only included the data for $z>-2$ in Fig.\ 
\ref{fig:gexil} to show this behaviour. As compared to the transverse
scaling function in the high temperature region we have a different
functional form, which is similar to that of the scaling function of the 
longitudinal susceptiblity in Fig.\ \ref{fig:fchi}. Like in the Ising case,
both functions have a peak at about the same $z$-value, which is at 
$z_p\approx 1.335$ in $f_{\chi}$, in accord with the value 1.33(5) found 
in Ref.\ \cite{Engels:2001bq}. As in Fig.\ \ref{fig:gexit} we find an early
approach to the asymptotic form given by Eq.\ (\ref{gxiasy}).

\section{The $H$-Dependence of $\xi_T$ at Fixed $T<T_c$}
\label{section:Hattlt0}
\begin{figure}[t]
\begin{center}
   \epsfig{bbllx=63,bblly=265,bburx=516,bbury=588,
       file=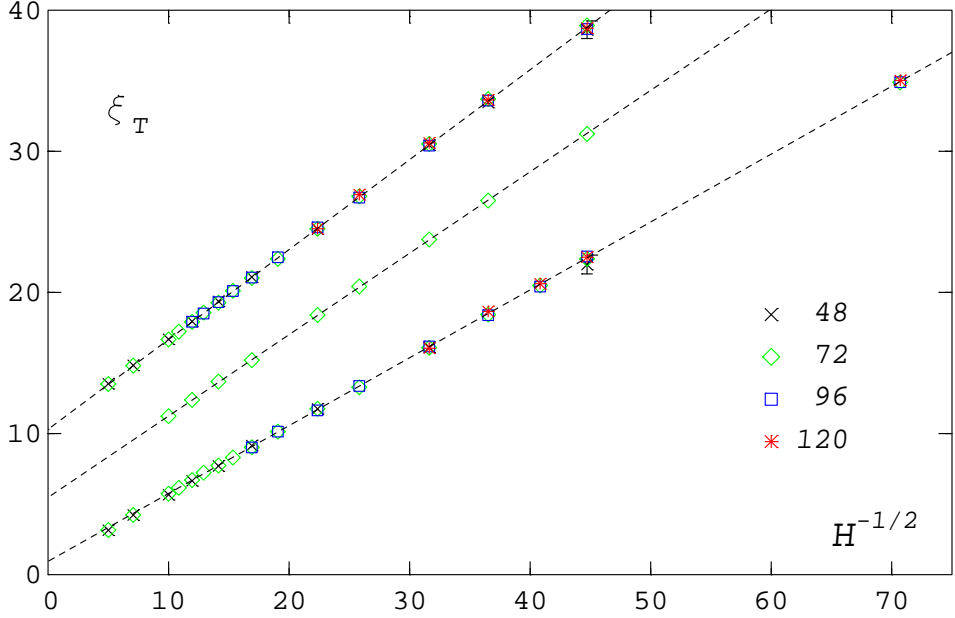, width=120mm}
\end{center}
\caption{The transverse correlation length as a function of $H^{-1/2}$
for $J=0.95,0.975$ and $1.0\,$. The lowest values are those of $J=0.95$,
the highest belong to $J=1.0\,$. For better visibility we have lifted 
the results for $J=0.975\,(1.0)$ by $5.0\,(10.0)$. The dashed lines
represent the straight line fits (\ref{xiline}) to the data.}
\label{fig:xitgold}
\end{figure}
Goldstone-mode effects are expected for all temperatures below $T_c$,
not only in the critical region, but also there. We have discussed already 
in the last section, that the shape of the scaling function for the 
transverse correlation length is explicitly determined in its asymptotic
part for $z\rightarrow -\infty$ by the predicted dependency of $\xi_T$
on $H^{-1/2}$ for $H\rightarrow 0$. We have tested this prediction by 
calculating $\xi_T$ at three fixed temperatures below $T_c$. The data
for $\xi_T$ are plotted versus $H^{-1/2}$ in Fig.\ \ref{fig:xitgold}
for $J=0.95,0.975$ and $J=1.0\:$. Actually, the same data points
from the larger lattices, apart from those for $J=1.0$, had already been
included in our scaling plot. We see from Fig.\ \ref{fig:xitgold} that the
data always follow a straight line
\be
\xi_T\;=\; x_0(J) H^{-1/2} + x_1(J)~,
\label{xiline}
\ee
where the slope is slightly increasing with increasing $J$: $x_0=0.481(2),
0.577(1)$ and 0.637(2) for the three $J$-values. The temperature
dependence of $x_0$ can be derived from the scaling function by comparing
its leading asymptotic behaviour to the $H$-dependence of $\xi_T$. If
\be
g_{\xi}^T(z) \;=\; g_q (-z)^q + \dots~,
\label{gtas}
\ee
with $g_q$ a constant, then the power $q$ must be $q=\beta\delta/2-\nu$
to lead to the required $H$-dependence for $\xi_T$. That implies 
\be
x_0\;=\; g_q (-\bar t)^q H_0^{1/2}~,
\label{x_0}
\ee
and since $q=0.1789$ is positive, we have an increase with increasing $J$.
We can check the $J$-dependence of the slopes $x_0$ by taking ratios
\be
{x_0(J_1) \over x_0 (J_2)} = \left( {-\bar t_1 \over -\bar t_2}\right)^q~.
\label{slope}
\ee
For $x_0(0.975)/ x_0 (0.95)=1.200(5)$ our formula predicts 1.195, that is
rather close to the fit value. For $x_0(1.0)/ x_0 (0.975)=1.104(4)$ we
get 1.088, a slightly lower result. On the other hand, the data for $J=1.0$
were just outside the scaling region and somewhat higher than the scaling
function. The small deviation to a higher ratio is therefore what one 
expects. 
\section{The $H$-Dependence of $\xi_{T,L}$ at Fixed $T>T_c$}
\label{section:Hattgt0}
\begin{figure}[t]
\begin{center}
   \epsfig{bbllx=95,bblly=265,bburx=484,bbury=588,
       file=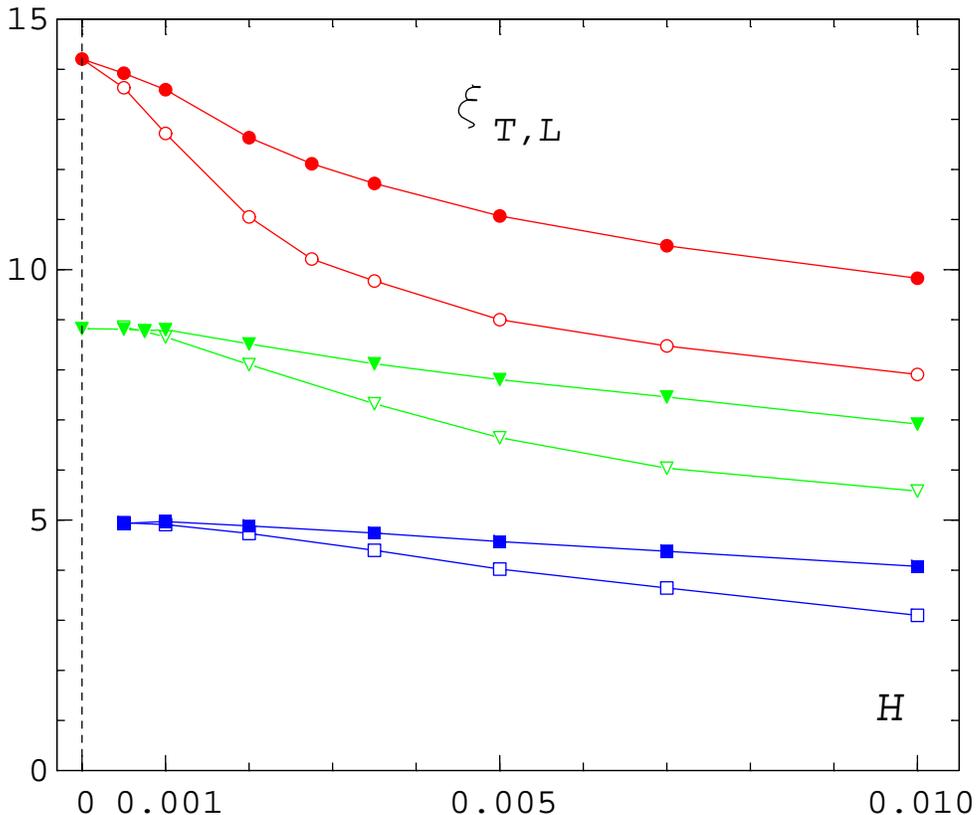, width=120mm}
\end{center}
\caption{The correlation lengths $\xi_T$ (filled symbols) and $\xi_L$
(empty symbols) as a function of $H$. The couplings are $J=0.92$ (circles), 
$J=0.91$ (triangles) and $J=0.90$ (squares). In order to disentangle the
different curves we have lifted the results for $J=0.92\,(0.91)$ by 5.0
(2.5). The data are connected by straight lines to guide the eye.}
\label{fig:xihight}
\end{figure}

\n At $H=0$ the correlation lengths in the symmetric phase $(T>T_c)$ are
equal. We have shown the $t$-dependence in Fig.\ \ref{fig:xiat}.
With increasing $H$, however, they differ and $\xi_T$ is always bigger than
$\xi_L$. In Fig.\ \ref{fig:xihight} we show the two correlation lengths as
a function of $H$ for three different fixed couplings or temperatures:
$J=0.92,0.91$, and 0.90. The lattice sizes were $L=48$ and 72. Essentially
no finite size effect was found. As expected, the correlation lengths
decrease with increasing field from their value at $H=0$. The reduction 
itself is diminishing with increasing temperatures (decreasing couplings
$J=1/T$), the curves become flatter and the difference between the two
correlation lengths disappears. This behaviour can be understood from the 
asymptotic expansions of the scaling functions of the correlation lengths.
In the symmetric phase the correlation lengths are even functions of $h$,
so that $\xi(h)=\xi(-h)$ (we omit the indices $T$ and $L$ for this
discussion). The scaling function $g_{\xi}(z)$ must then have an asymptotic 
expansion for $z\rightarrow \infty\: (h\rightarrow 0, \bar t$ fixed) of the
form
\be
g_{\xi}(z) \;=\; z^{-\nu} \sum_{n=0}^{\infty} g_n z^{-2\Delta n}~,
\label{gasex}
\ee 
\noindent where the $g_n$ are constants, $g_0=\xi^+ B^{\nu/\beta}$, and 
$\Delta=\beta\delta$. This translates into the following $h$-expansion
for the correlation length at fixed $\bar t$ and small $h$
\be
\xi \;=\; (\bar t)^{-\nu}  \sum_{n=0}^{\infty} g_n\cdot 
(\bar t)^{-2\Delta n} h^{2n}~.
\label{xiasex}
\ee  
For very small magnetic fields $h$ we have
\be
\xi \;=\; g_0 (\bar t)^{-\nu} + g_1 (\bar t)^{-\nu-2\Delta} h^2+ \dots~, 
\label{firstt}
\ee 
and with increasing $\bar t$ the $h$-dependence will disappear with a 
factor $\sim (\bar t)^{-2\Delta}$ compared to the value at $h=0$.

\begin{figure}[t]
\begin{center}
   \epsfig{bbllx=63,bblly=280,bburx=516,bbury=563,
       file=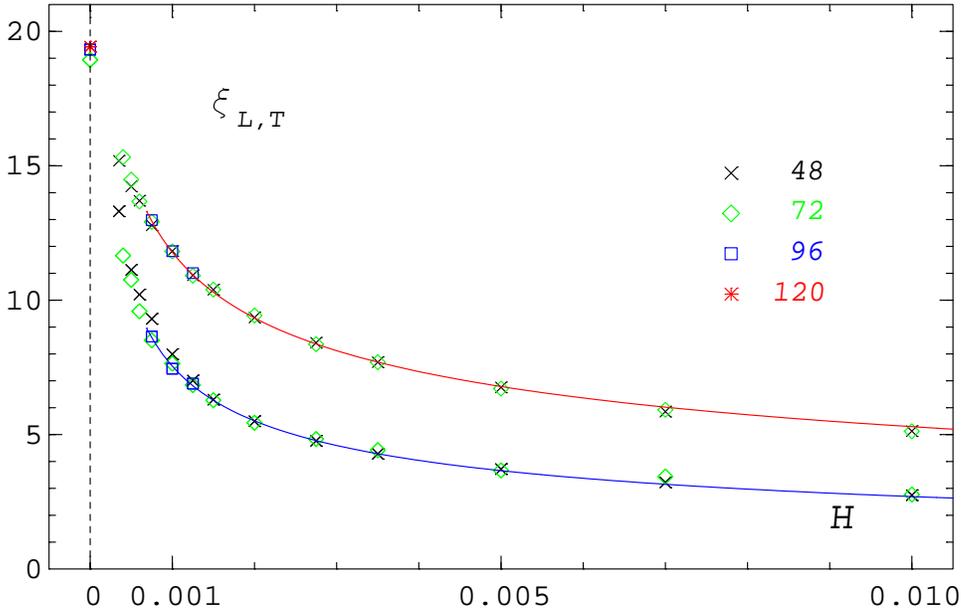, width=120mm}
\end{center}
\caption{The correlation lengths $\xi_T$ (upper set of data) and $\xi_L$
(lower set) as a function of $H$ for the coupling $J=0.93$. The lines 
are fits to Eq.\ (\ref{hdep}) in the interval [0.0006,0.005].}
\label{fig:xi93}
\end{figure}
In the neighbourhood of the critical point, where $|z|$ is small (but not 
$h$), one can expand $g_{\xi}(z)$ in powers of $z$ and obtains
\be
\xi \;=\; h^{-\nu_c} g_{\xi}(z) \;=\;  h^{-\nu_c} \left(
g_0^c + g_1^c z +g_2^c z^2 +\dots\right)~,
\label{Taylor}
\ee
with $g_0^c = \xi^c (B^c)^{\nu/\beta}$. That leads to the following 
$h$-dependence of $\xi$
\be
\xi \;=\; h^{-\nu_c} \left( g_0^c + g_1^c {\bar t} h^{-1/\Delta} +
 g_2^c (\bar t)^2 h^{-2/\Delta} + \dots \right)~,
\label{hdep}
\ee
when the temperature is near to the critical point. In Fig.\ \ref{fig:xi93}
we show as an example the correlation lengths for $J=0.93$, close
to $J_c=0.93590\:$. We have made fits in the $H$-interval [0.0006,0.005], 
using only the first two terms in Eq.\ (\ref{hdep}). Obviously these
simple fits describe the data quite well in a large range.   
\section{Summary and Conclusions}
\label{section:conclusion}
In this paper we have investigated the correlation lengths of the 
three-dimensional $O(4)$ model with Monte Carlo simulations. One of the
main objectives was to find the functional dependence of the correlation
lengths on the external field at fixed temperature, a second to understand
the interplay of the critical behaviour and the effects induced by the
massless Goldstone-modes. A main result of our work is the calculation
of the scaling function of the transverse correlation length and that of 
the longitudinal correlation length for $T\ge T_c$. In the low temperature
phase we were, however, unable to reliably estimate $\xi_L$. The reason
for that is the rich spectrum of states which contribute to the 
longitudinal correlators below the critical point. In the simpler Ising
model \cite{Engels:2002fi}, where only one correlation length exists, 
the influence of higher states \cite{Caselle:1999tm}
complicates the determination of the correlation length below $T_c$, 
but in contrast to the $O(4)$ model it is still possible there.
 
 As we could show, the scaling functions do not only describe the critical
behaviours of the correlation lengths, but encompass as well the predicted
Goldstone effects. Indeed, in consequence of the relation $\xi_T^2\sim\chi_T$
and the equation $\chi_T\,=\,M/H$ for the fluctuation of the transverse 
spin components, the prediction $\xi_T\sim H^{-1/2}$ emerges below $T_c$
and leads to a similar functional form of the scaling functions of $\xi_T$
and that of the magnetization. In the high temperature phase a similar
correspondence exists between the scaling functions of $\xi_L$ and that of
the longitudinal susceptibility. There both functions have a peak at the
same position, a behaviour which was found also in the Ising model
\cite{Engels:2002fi}.

 In preparation of the scaling functions we have determined several critical
amplitudes of the correlation lengths: $\xi_T^c,\, \xi_L^c$ and $\xi^+$ and 
in addition also $B^c$ and $C^+$. As a byproduct we have found the
critical exponent $\delta\,=\, 4.824(9)$. The critical amplitudes allowed
us to directly determine the universal amplitude ratios $R_{\chi},\,Q_c,\,
Q_2^{L,T}$ and $U_{\xi}^c\,=\,1.99(1)\,$. The latter result is
most remarkable, because it confirms a relation between the longitudinal
and transverse correlation functions, which was conjectured to hold below
$T_c$, but seems to be valid also at $T_c$.
\vskip 0.2truecm
\noindent{\Large{\bf Acknowledgments}}

\n We thank Sven Holtmann and Thomas Schulze for their constant interest and
assistance during the course of the work. Our work was supported by the Deutsche 
Forschungs\-ge\-meinschaft under Grant No.\ FOR 339/2-1.

\clearpage


\begin{thebibliography}{99}

\bibitem{Fisher:1985}
M.~E.~Fisher and V.~Privman,
Phys.\ Rev.\ B {\bf 32} (1985) 447.

\bibitem{Patashinskii:1973}
A.~L.~Patashinskii and V.~L.~Pokrovskii,
Zh.\ Eksp.\ Teor.\ Fiz.\ {\bf 64} (1973) 1445 [Sov.\ Phys.\ JETP
{\bf 37} (1974) 733].

\bibitem{Zinn-Justin:1996}
J.~Zinn-Justin, {\em Quantum Field Theory and Critical Phenomena},
Clarendon Press, Oxford, 3rd Edition 1996.

\bibitem{Anishetty:1995kj}
R.~Anishetty, R.~Basu, N.~D.~Hari Dass and H.~S.~Sharatchandra,
Int.\ J.\ Mod.\ Phys.\ A {\bf 14} (1999) 3467
[hep-th/9502003].

\bibitem{Pisarski:ms}
R.~D.~Pisarski and F.~Wilczek,
Phys.\ Rev.\ D {\bf 29} (1984) 338.

\bibitem{Wilczek:1992sf}
F.~Wilczek,
Int.\ J.\ Mod.\ Phys.\ A {\bf 7} (1992) 3911
[Erratum-ibid.\ A {\bf 7} (1992) 6951].

\bibitem{Rajagopal:1992qz}
K.~Rajagopal and F.~Wilczek,
Nucl.\ Phys.\ B {\bf 399} (1993) 395
[hep-ph/9210253].

\bibitem{Engels:2001bq}
J.~Engels, S.~Holtmann, T.~Mendes and T.~Schulze,
Phys.\ Lett.\ B {\bf 514} (2001) 299
[hep-lat/0105028].

\bibitem{Toldin:2003hq}
F.~P.~Toldin, A.~Pelissetto and E.~Vicari,
JHEP {\bf 0307} (2003) 029
[hep-ph/0305264].

\bibitem{Toussaint:1996qr}
D.~Toussaint,
Phys.\ Rev.\ D {\bf 55} (1997) 362
[hep-lat/9607084].

\bibitem{Engels:1999wf}
J.~Engels and T.~Mendes,
Nucl.\ Phys.\ B {\bf 572} (2000) 289
[hep-lat/9911028].

\bibitem{Privman:1991}
V.~Privman, P.~C.~Hohenberg and A.~Aharony,
in {\em Phase Transitions and Critical Phenomena}, vol. 14,
edited by C.~Domb and J.~L.~Lebowitz (Academic Press, New York, 1991).

\bibitem{Vaks:1968} 
See e.g. V.~G.~Vaks, A.~I.~Larkin and S.~A.~Pikin, 
Sov.\ Phys.\ JETP {\bf 26} (1968) 647.

\bibitem{Wallace:1975} 
D.~J.~ Wallace and R.~K.~P.~Zia,
Phys.\ Rev.\ B {\bf 12} (1975) 5340.

\bibitem{Engels:2002fi}
J.~Engels, L.~Fromme and M.~Seniuch,
Nucl.\ Phys.\ B {\bf 655} (2003) 277
[cond-mat/0209492].

\bibitem{Pelissetto:2000ek}
A.~Pelissetto and E.~Vicari,
\PRep {\bf 368} (2002) 549
[cond-mat/0012164] .

\bibitem{Widom:1965}
B.~Widom,
J.\ Chem.\ Phys. {\bf 43} (1965) 3898.

\bibitem{Griffiths:1967}
R.~B.~Griffiths,
Phys.\ Rev.\ {\bf 158} (1967) 176.

\bibitem{Wolff:1988uh}
U.~Wolff,
Phys.\ Rev.\ Lett.\  {\bf 62} (1989) 361.

\bibitem{Dimitrovic:yd}
I.~Dimitrovic, P.~Hasenfratz, J.~Nager and F.~Niedermayer,
Nucl.\ Phys.\ B {\bf 350} (1991) 893.

\bibitem{Oevers:1996dt}
M.~Oevers, Diploma thesis, Universit\"at Bielefeld, 1996.

\bibitem{Caselle:1997hs}
M.~Caselle and M.~Hasenbusch,
J.\ Phys.\ A {\bf 30} (1997) 4963
[hep-lat/9701007].

\bibitem{Hasenbusch:2000ph}
M.~Hasenbusch,
J.\ Phys.\ A {\bf 34} (2001) 8221
[cond-mat/0010463].

\bibitem{Cucchieri:2002hu}
A.~Cucchieri, J.~Engels, S.~Holtmann, T.~Mendes and T.~Schulze,
J.\ Phys.\ A {\bf 35} (2002) 6517
[cond-mat/0202017].

\bibitem{Oku:1979ht}
M.~Oku and Y.~Okabe,
Prog.\ Theor.\ Phys.\  {\bf 61} (1979) 443.

\bibitem{Caselle:1999tm}
M.~Caselle, M.~Hasenbusch and P.~Provero,
Nucl.\ Phys.\ B {\bf 556} (1999) 575
[hep-lat/9903011].

\end{thebibliography}
\end{document}